\documentclass[%
 reprint,
 amsmath,amssymb,
 aps, longbibliography
]{revtex4-1}

\usepackage{graphicx}
\usepackage{dcolumn}
\usepackage{bm}
\usepackage{tipa}
\usepackage{upgreek}

\usepackage{xcolor, soul}
\sethlcolor{yellow}

\begin{document}

\title{Emergent collective motion of self-propelled condensate droplets}


\author{Marcus Lin$^{1}$}
\author{Philseok Kim$^{2}$}
\author{Sankara Arunachalam$^{1}$}
\author{Rifan Hardian$^{1}$}
\author{Solomon Adera$^{2}$}
\author{Joanna Aizenberg$^{2, 3}$}
	\email{jaiz@seas.harvard.edu}
\author{Xi Yao$^{2, 4}$}
	\email{xi.yao@cityu.edu.hk}
\author{Dan Daniel$^{1}$}
    \email{danield@kaust.edu.sa}

\affiliation{$^{1}$Division of Physical Sciences and Engineering, King Abdullah University of Science and Technology (KAUST), Thuwal 23955-6900, Saudi Arabia}
\affiliation{$^{2}$John A. Paulson School of Engineering and Applied Sciences, Harvard University, Cambridge, MA 02138, USA}
\affiliation{$^{3}$Department  of  Chemistry  and  Chemical  Biology,  Harvard University, Cambridge, MA 02138, USA}
\affiliation{$^{4}$Department of Biomedical Sciences, City University of Hong Kong, Hong Kong, China}

\begin{abstract}

Recently, there is much interest in droplet condensation on soft or liquid/liquid-like substrates. Droplets can deform soft and liquid interfaces resulting in a wealth of phenomena not observed on hard, solid surfaces (e.g., increased nucleation, inter-droplet attraction). Here, we describe a unique complex collective motion of condensate water droplets that emerges spontaneously when a solid substrate is covered with a thin oil film. Droplets move first in a serpentine, self-avoiding fashion before transitioning to circular motions. We show that this self-propulsion (with speeds in the 0.1--1 mm s$^{-1}$ range) is fuelled by the interfacial energy release upon merging with newly condensed but much smaller droplets. The resultant collective motion spans multiple length scales from sub-millimetre to several centimetres, with potentially important heat-transfer and water-harvesting applications. 
\end{abstract}
\maketitle



With its long and illustrious history, the study of condensate droplets (sometimes called breath figures) has captured the imaginations of many scientists including Lord Rayleigh and C.V. Raman \cite{lord1911breath, raman1921colours, knobler1988growth, beysens2006dew, beysens2022physics}. Optimizing the condensation process, such as promoting dropwise condensation (as opposed to filmwise condensation) and maximizing droplet mobility \cite{boreyko2009self}, has important heat-transfer and water-harvesting applications \cite{rose2002dropwise}. The ability to control the size and spatial locations of condensate droplets is also useful when producing micro-/nano-structured materials \cite{zhang2015breath} and structural colours \cite{raman1921colours, goodling2019colouration}.  
 
More recently, there is much interest in understanding droplet condensation on lubricated surfaces \cite{anand2012enhanced, anand2015droplets, xiao2013immersion, sun2019microdroplet, sun2023marangoni} for enhanced heat-transfer and water-collection applications \cite{anand2012enhanced, park2016condensation, adera2021enhanced, tripathy2021ultrathin}. Water droplets are highly mobile on lubricated surfaces \cite{lafuma2011slippery, wong2011bioinspired, smith2013droplet}, because the presence of a stable lubricant film (typically silicone or fluorinated oils) prevents direct contact of the droplet with the underlying solid substrate and hence eliminates contact-line pinning \cite{daniel2017oleoplaning, keiser2020universality}. The lubricant also tends to wrap around droplets forming menisci or wetting ridges around them, which can overlap and give rise to inter-droplet capillary attraction (akin to the Cheerios effect) \cite{vella2005cheerios}. In the absence of contact-line pinning, the droplets will therefore move towards each other before coalescing into one larger droplet \cite{sun2019microdroplet, jiang2019directional, vella2005cheerios}. This increased rate of droplet coalescence frees up new areas for renucleation of new droplets and therefore increases heat transfer rate. Previous reports of capillary attraction is confined to neighbouring droplets, and droplet displacement is limited to a millimetre or less with no long-range collective motion \cite{aizenberg2000patterned, anand2012enhanced, sun2019microdroplet, jiang2019directional}. 

In contrast, here we report the spontaneous emergence of complex collective motion of condensate droplets whose displacements span several centimetres or more, i.e., multiple times the droplet diameters. Droplets ranging from tens of microns to several millimetres in size move laterally in a serpentine, self-avoiding fashion, i.e., droplets preferentially avoid their own and each other's previous paths. At a later stage when the \textit{local} lubricant film is depleted, the serpentine motion transitions to circular motions. As the lubricant is continually being redistributed spatially by the sweeping droplet motions, the droplets continually switch between serpentine and circular motions in a highly collective fashion---a classic example of emergence where complexity arises from simple interactions of its individual parts \cite{sep-properties-emergent, anderson1972more}. The observed lateral speeds $U_{\text{L}}$ (in the 0.1--1 mm s$^{-1}$ range) is fuelled by the interfacial energy release upon merging with newly condensed but much smaller droplets. A similar physical argument---the conversion of interfacial to kinetic energy---was proposed for jumping droplets on superhydrophobic surfaces \cite{boreyko2009self, enright2014coalescing, lecointre2019ballistics, mouterde2017merging}, except the self-propulsion on superhydrophobic surfaces occurs in the vertical, out-of-plane direction (instead of lateral, in-plane motion). The phenomenon described here is an interesting example of active matter \cite{sanchez2012spontaneous, vicsek2012collective, michelin2023self} whereby autonomous droplet motions are fuelled by the energy release from condensation \cite{steyer1992spontaneous}, as opposed to the more typical chemical reactions or Marangoni effects  \cite{michelin2023self, izri2014self, meredith2020predator, wen2019vapor, liu2019reconfigurable, paxton2006chemical}.

\begin{figure}[!htb]
\centering
\includegraphics[scale=1]{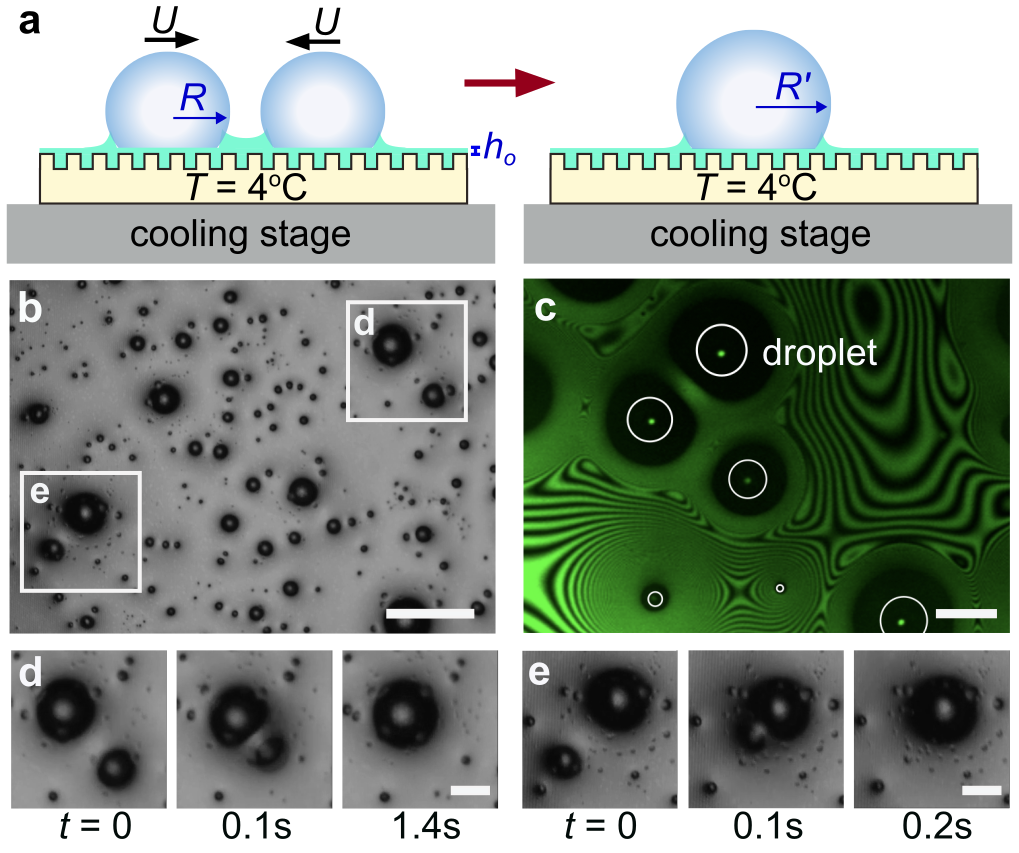}
\caption{\label{fig:attraction} \textit{Initial inter-droplet attraction} (a) Inter-droplet attraction and coalescence on a cooled lubricated surface ($T$ = 4$^{\circ}$C). (b) Coalescence and subsequent recondensation results in polydisperse droplet sizes. Scale bar is 2 mm. (c) Non-axisymmetric menisci (dark regions around each droplet) as observed using reflection interference contrast microscopy. Scale bar is 0.1 mm. (d)--(e) Time series for the two boxed regions in (b). Scale bars are 0.15 mm.}
\end{figure}

\begin{figure}[!htb]
\centering
\includegraphics[scale=1]{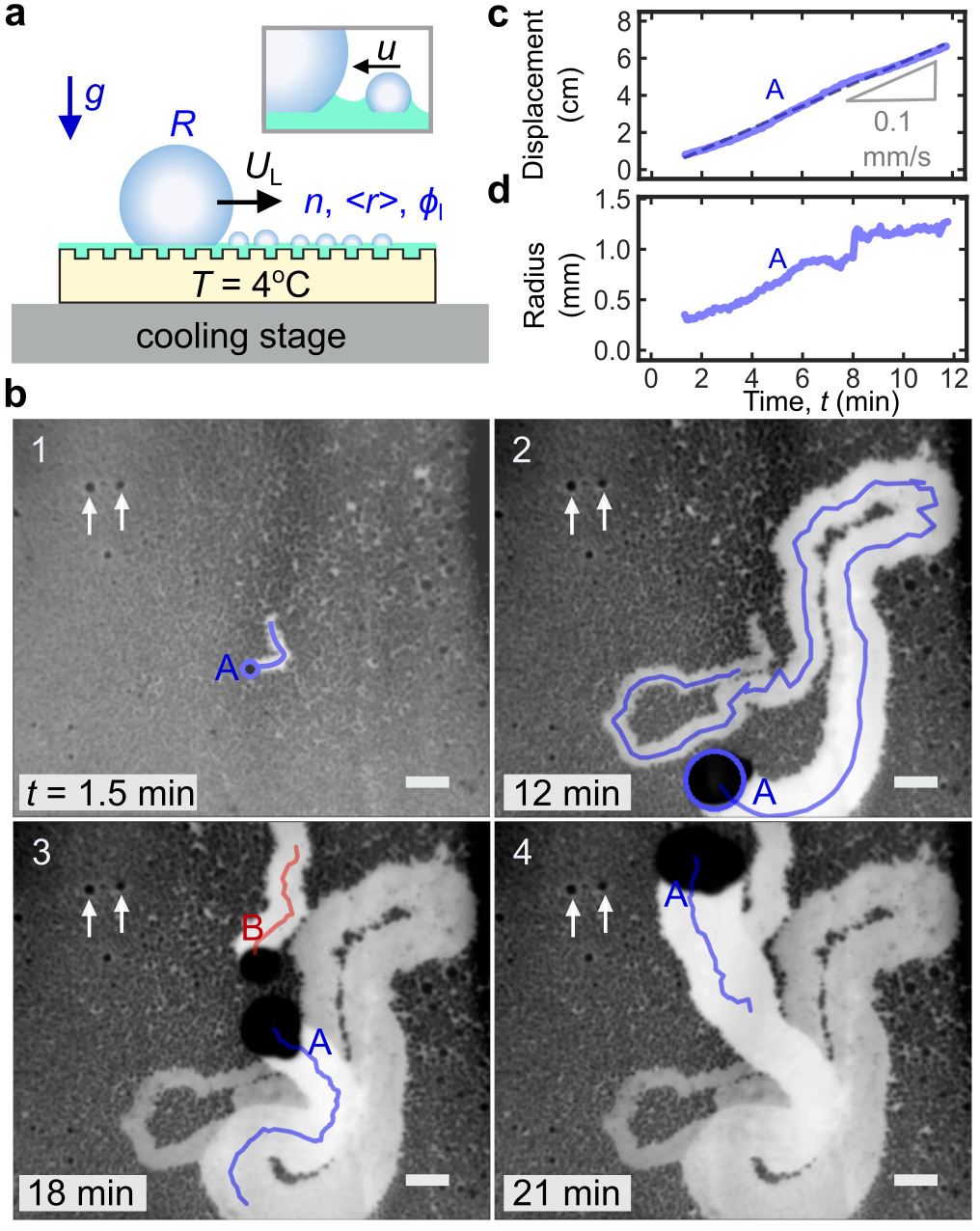}
\caption{\label{fig:serpentine} \textit{Self-avoiding motion on cooled surface.} (a) Schematic and (b) IR imaging (2 mm scale bars) showing the self-propelled serpentine motion of condensate droplets on a cooled lubricated micropillar surface. Droplet size and position are tracked automatically using Hough transform \cite{opencv_library}. Increase in (c) displacement and (d) radius with time for droplet A for the first 12 min of motion (panel 1 to 2 in (b)). $t$ = 0 min corresponds to the start of droplet A motion which occurs after 27 minutes of cooling.}
\end{figure}

We start by cooling a micropillar surface (hexagonal array with diameter $d$ = 2 $\upmu$m, pitch $p$ = 10 $\upmu$m, and height $h_{p}$ = 9 $\upmu$m) lubricated with a thin film of fluorinated oil (overlayer thickness $h_{o}$ = 5 $\upmu$m, viscosity $\eta = 53$ mPa.s) to a temperature of $T$ = 4$^{\circ}$C. Throughout the experiments described here, the ambient temperature was kept at 22$^{\circ}$C and the room humidity between 40 and 60$\%$. Within the first minute, water started to condense on the surface, and we observed capillary attraction between neighbouring droplets (sub-millimetric and \textit{similarly} sized). This occurs when their wetting ridges overlap resulting in a non-axisymmetric meniscus around each droplet and hence a net attractive force akin to the Cheerios effect [Figs.~\ref{fig:attraction}(a)--\ref{fig:attraction}(b)] \cite{vella2005cheerios, sun2019microdroplet, jiang2019directional}. The deformation to the lubricant interface can be visualized by shining monochromatic light with wavelength $\lambda$ = 561 nm from below, a technique known as reflection interference contrast microscopy [Fig.~\ref{fig:attraction}(c)] \cite{daniel2017oleoplaning, limozin2009quantitative}; the bright and dark fringes correspond to different lubricant heights, with a step height $|\Delta h| = \lambda/4 n_{o} \approx$ 100 nm between one bright fringe and a neighbouring dark fringe, where $n_{o}$ = 1.3 is the refractive index of the lubricant. We found that the size of wetting ridge (dark region around the droplet outline) is comparable to that of the droplet (See Supporting Fig.~S1 for detailed interpretation of Fig.~\ref{fig:attraction}(c)). Hence, droplets that are several diameters apart can move towards each other before coalescing into one larger droplet (Figs.~\ref{fig:attraction}(d)--\ref{fig:attraction}(e); see also Supporting Videos S1). 

The details of this inter-droplet attraction have been described by many others \cite{sun2019microdroplet, jiang2019directional, boreyko2014air}. Here, we would like to emphasize that this process naturally leads to polydisperse droplet sizes [Fig.~\ref{fig:attraction}(b)]: coalescence results in larger droplets, while at the same time clears up new areas for recondensation and the formation of much smaller droplets. This polydispersity is key to understanding the mechanism behind the collective droplet motion described below. 

As time progresses (after 27 minutes of cooling), one droplet becomes much larger than its neighbours and starts moving in a serpentine, self-avoiding fashion, as imaged using an infra-red (IR) camera (Figs.~\ref{fig:serpentine}(a)--\ref{fig:serpentine}(b); Supporting Videos S2 and S3). Unlike the droplet propulsion described previously which is limited to sub-millimetric distances, the serpentine motion here can cover much larger distances. For example, droplet A in Fig.~\ref{fig:serpentine}(b)-1 traversed a distance of 6 cm within 12 minutes with an average lateral velocity of $U_{\text{L}}$ = 0.1 mm s$^{-1}$ [Fig.~\ref{fig:serpentine}(c)]; as the droplet swept through its much smaller neighbours and incorporated their volumes, its radius $R$ more than triples from less than 0.4 mm to 1.3 mm [Fig.~\ref{fig:serpentine}(d)]. The serpentine motion is therefore an important way by which droplets grow (cf. stationary droplets indicated by arrows in Fig.~\ref{fig:serpentine}(b)), which further increases the droplet polydispersity. 

The self-propelled droplets are also self-avoiding: they preferentially avoid their own and each other's paths (See how Droplet B avoids the path of Droplet A in Fig.~\ref{fig:serpentine}(b)-3). The droplets can only intersect their own and other's paths once there is sufficient recondensation on their previous paths (which consequently turn from white to gray). While the self-propelled droplets avoid each other's paths, they can collide into each other, coalesce into a larger droplet which starts its own serpentine motion [Fig.~\ref{fig:serpentine}(b)-4]. Note that we positioned our surface horizontally, and gravity-assisted shedding is not an important factor unlike in previous studies \cite{anand2012enhanced}. Gravity plays little to no role as long as $R \ll l_{c}$, where $l_{c}$ = $\sqrt{\gamma/\rho g}$ = 2.7 mm is the capillary length. 

The origin of this self-propelled serpentine motion is similar to the interdroplet attraction described previously except that the capillary attraction is between one large droplet and multiple smaller droplets. The attraction occurs once again when their wetting ridges overlap (Fig.~\ref{fig:serpentine}(a) inset; Supporting Fig.~S2). The capillary force arising from the non-axisymmetric menisci is notoriously difficult to model mathematically, and analytic solutions only exist when droplets are far apart \cite{vella2005cheerios, kralchevsky2000capillary, chan1981interaction}.  However, we can overcome this difficulty and derive a simple expression for $U_{\text{L}}$ by using energy balance arguments outlined below. 

As the large droplet of radius $R$ gobbles up its neighbouring droplets of much smaller mean radius $\langle r \rangle$ and number density (per unit area) $n$, the interfacial energy of the smaller droplets $E_{\gamma}$ is released at a rate of 
\begin{equation} \label{eq:dEdt}
\begin{split}
\frac{\mathrm{d}E_{\gamma}}{\mathrm{d}t}  &= (2RU_{\text{L}})(2n\pi \langle r^{2} \rangle)\gamma \\
 								 &=  4 n\pi \langle r^{2} \rangle RU_{\text{L}} \gamma \\
 								 &= 4 \phi_{l} RU_{\text{L}} \gamma
\end{split}
\end{equation}
where $\gamma$ is the droplet's surface tension, $\phi_{l}$ = $n\pi \langle r \rangle^{2}$ is the \textit{local} fractional surface coverage of the smaller droplets approximated here as hemispheres with surface areas $2\pi \langle r^{2} \rangle$ and projected areas $\pi \langle r^{2} \rangle$. Note that in our experiments, the lubricant oil (silicone or fluorinated) tends to wrap around the water droplet. Hence, $\gamma$ is strictly speaking the effective surface tension whose magnitude varies between 60 and 70 mN m$^{-1}$ depending on the choice of lubricant (See Table III in \citeauthor{kreder2018film} 2018 \cite{kreder2018film}), but not too far from the 72 mN m$^{-1}$ value for pure water droplet. 

A fraction $\alpha < 1$ of this interfacial energy is converted to translational kinetic energy of the large droplet, which in turn must be balanced by viscous dissipation. Previously, we and others showed that the viscous force for a droplet moving on lubricated surface is given by $F_{\eta}$ = 16 $\gamma R$ Ca$^{2/3}$, where Ca = $\eta U_{\text{L}}/\gamma$ is the capillary number \cite{daniel2017oleoplaning, keiser2020universality}. Hence, we expect that 
\begin{equation} \label{eq:balance}
\begin{split}
F_{\eta} \, U_{\text{L}}                                &=  4\alpha \phi_{l} RU_{\text{L}} \gamma  \\
										   U_{\text{L}} &= \frac{\gamma}{\eta} \left( \frac{\alpha \phi_{l}}{4 }\right)^{3/2} \\
\end{split}
\end{equation}
. The droplet motion is therefore fuelled by the coalescence with the surrounding water droplets of \textit{local} area coverage $\phi_{l}$, with $\gamma/\eta$ being the characteristic visco-capillary speed which is temperature dependent (See Supporting Figs.~S3--S5 for a rigorous derivation and justification). Self-avoiding motion arises because its previous path contains little to no water content to fuel self-propulsion (i.e., $\phi_{l} = 0$ and hence appears white), and droplets continually seek areas with \textit{locally} higher $\phi_{l}$ (darker regions) [Supporting Fig.~S4 and Supporting Video S4]. 

Interestingly, our model predicts that $U_{\text{L}}$ is independent of droplet radius $R$, which is consistent with the experimental results in Figs.~\ref{fig:serpentine}(c)--\ref{fig:serpentine}(d): the droplet moves at relatively constant speed despite tripling its size. This is in sharp contrast to the vertical velocity of jumping droplets $U_{\text{V}} \sim \sqrt{\gamma/\rho R}$ which is highly dependent on $R$. In our model, $U_{\text{L}}$ is also insensitive to the exact dimension of the small droplet $\langle r \rangle$ and depends only on its \textit{local} surface coverage $\phi_{l}$; the prediction that $U_{\text{L}}$ increases with $\phi_{l}$ is borne out \textit{qualitatively} in experiments (Supporting Fig.~S4). Note that the different grayscale values are not due to temperature variations, but changes in material emissivity when water condenses on the silicon substrate: the more water and the higher $\phi_{l}$ is, the darker the pixel appears (See Supporting Information `Materials and Methods' section). However, with IR imaging in Fig.~\ref{fig:serpentine}(b) (and Supporting Fig.~S4), $\phi_{l}$ value can only be deduced \textit{qualitatively}.

\begin{figure}[!htb]
\centering
\includegraphics[scale=1]{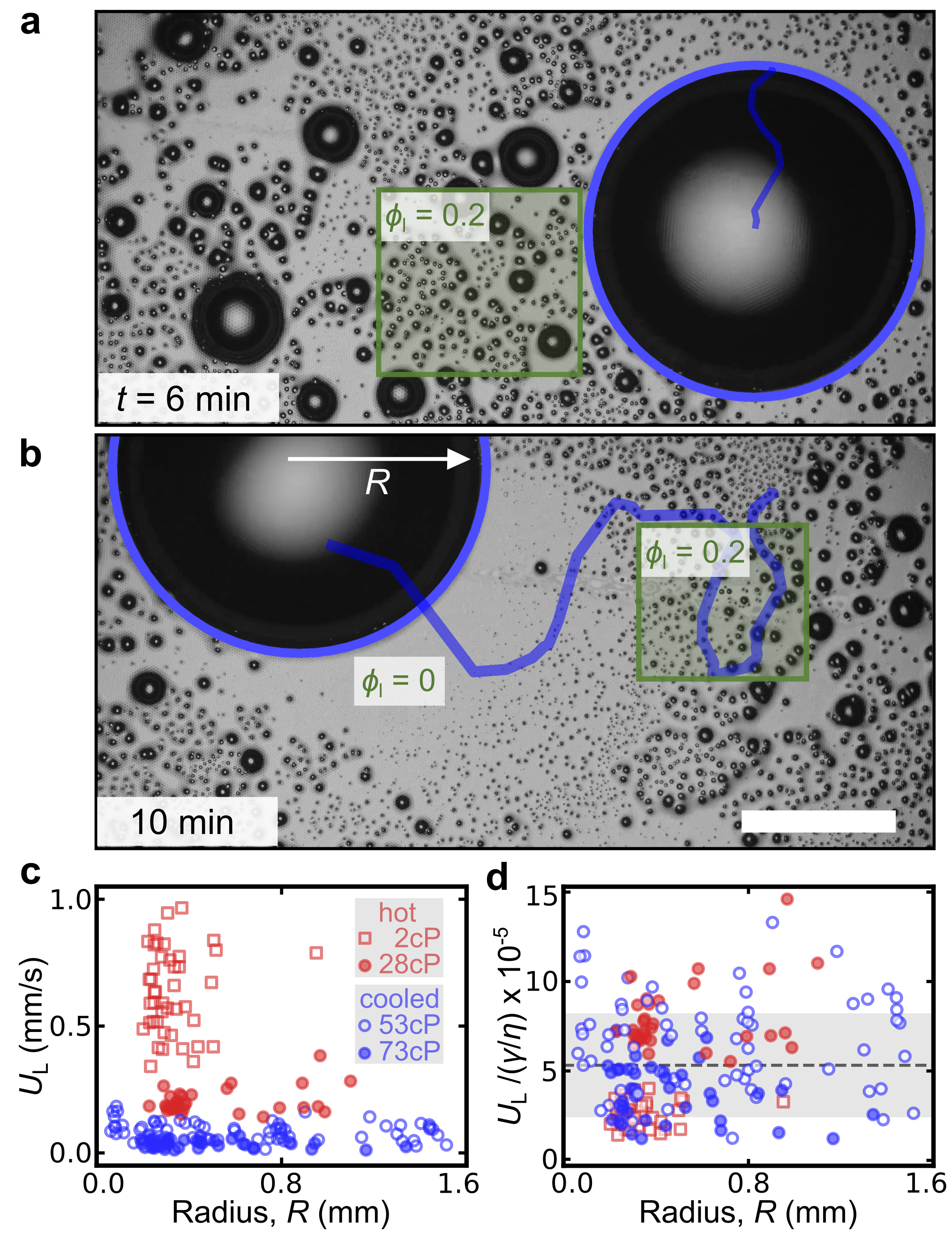}
\caption{\label{fig:serpentine_zoomed} \textit{Zoomed-in view of serpentine motion on cooled surface}. (a)--(b) Droplet with radius $R$ sweeping through neighbouring droplets of much smaller mean radius $\langle r \rangle$ and \textit{local} area coverage $\phi_{l}$. Scale bar is 1 mm. $t$ = 0 min corresponds to start of the droplet serpentine motion. (c) Velocity $ U_{\text{L}}$ and (d) normalised velocity $U_{\text{L}}/(\gamma/\eta)$ as a function of droplet radius $R$ on lubricated surfaces with different viscosities. Circles and squares represent fluorinated and silicone oils, respectively. We used nanotextured and micropillar surfaces for hot and cooled experiments, respectively.}
\end{figure}

To accurately quantify $\phi_{l}$, we imaged the droplets using a microscope objective and a high resolution camera as shown in Figs.~\ref{fig:serpentine_zoomed}(a)--\ref{fig:serpentine_zoomed}(b) and Supporting Video S5. We observed a large droplet ($R$ = 1.2 mm, circled blue) moving in a serpentine fashion, sweeping through much smaller droplets with $\langle r \rangle$ = 18 $\upmu$m and local  $\phi_{l} = 0.2$ [boxed green in Fig.~\ref{fig:serpentine_zoomed}(a)]. Immediately after an area is swept through, $\phi_{l} =$ 0 but its value returns to its original magnitude within minutes of recondensation [Fig.~\ref{fig:serpentine_zoomed}(b)]. Experimentally, we found that $\phi_{l}$ remains relatively constant at 0.15--0.30 over 4 hours of continuous cooling; this saturation value is significantly lower than 0.55 typically encountered on solid surfaces \cite{beysens2006dew} and arises due to a complex interplay between the sweeping motion of the large droplets and the recondensation rates, which in turn depends on surface subcooling and relative humidity [Supporting Figs.~S6--S8].     

\begin{figure}[!htb]
\centering
\includegraphics[scale=1]{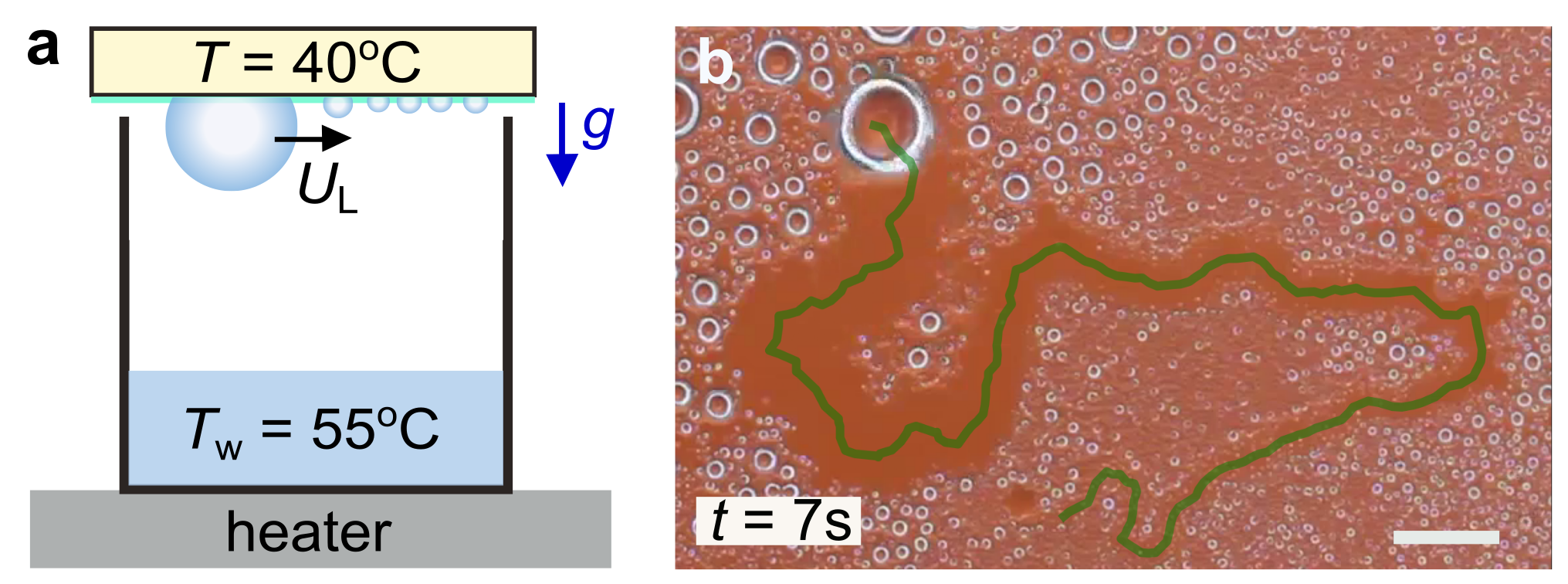}
\caption{\label{fig:serpentine_B} \textit{Self-avoiding motion of hot vapour condensate.} (a) Schematic and (b) photograph (0.5 mm scale bar) showing similar self-avoiding droplet motion from hot vapour condensing on lubricated nanotextured surface.}
\end{figure}

We measured $U_{L}$ for droplets of various sizes $R$ = 0.075--1.5 mm on two surfaces with different fluorinated oils cooled to $T$ = 4$^{\circ}$C [blue open and filled markers in Fig.~\ref{fig:serpentine_zoomed}(c)]. We found that $U_{L}$ remains constant at 0.09$\pm$0.03, 0.04$\pm$0.01 mm s$^{-1}$ for the two cases ($\eta$ = 53, 73 mPa.s or cP, respectively) over 4 hours of continuous cooling; more importantly, $U_{L}$ is independent of droplet size $R$ as predicted by Eq.~(\ref{eq:balance}). When hot vapour from a heated beaker of water (water temperature $T_{w}$ = 55$^{\circ}$C) condenses on a lubricated nanotextured surface (substrate temperature raised to $T$ = 40$^{\circ}$C), the condensate droplets also perform serpentine motion [Fig.~\ref{fig:serpentine_B} and Supporting Video S6]. The resulting propulsion speeds $U_{L} = 0.20\pm0.05$ and $0.63\pm0.17$ mm s$^{-1}$ are similarly independent of $R$ [red markers in Fig.~\ref{fig:serpentine_zoomed}(c)], but their magnitudes are higher than for cooled substrate because of the lower lubricant viscosity attained ($\eta$ = 28 and 2 mPa s$^{-1}$ for heated fluorinated and silicone oils, respectively). 

When scaled by the viscocapillary speed $\gamma/\eta$, all the velocity data overlap with one another with a mean value of $U_{L}/(\gamma/\eta) = (5 \pm 3) \times 10^{-5}$ [Fig.~\ref{fig:serpentine_zoomed}(d)]; the scatter in the data can be attributed to variations in local lubricant film thickness and wetting ridge geometries, which can influence viscous dissipation and are not accounted for by our model. If we compare the  $U_{L}/(\gamma/\eta)$ value with predictions of Eq.~(\ref{eq:balance}) and use the fact that $\phi_{l} \approx 0.2$, we conclude that $\alpha \approx 0.03$, i.e., only about 3$\%$ of the interfacial energy is converted to translational motion, with the rest of the energy likely to be converted into internal flow inside the droplet. Similarly low $\alpha \sim 1\%$ was reported for jumping droplets on superhydrophobic surfaces \cite{enright2014coalescing, yan2019droplet}.
 
\begin{figure*}[!htb]
\centering
\includegraphics[scale=1.043]{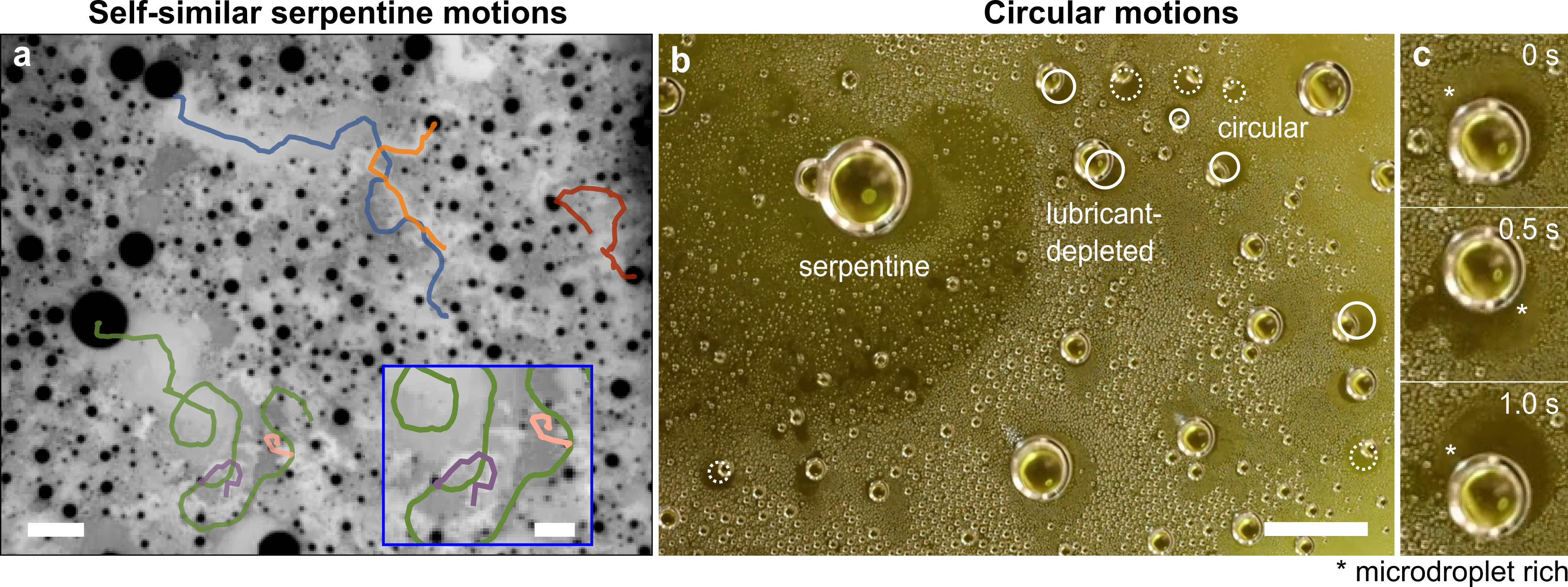}
\caption{\label{fig:complex} \textit{Self-similar serpentine and circular motions of condensate droplets.} (a) IR imaging of self-similar serpentine droplet motions on cooled substrate. Scale bars are 2 mm and 1 mm for main figure and inset. (b) Multiple circular motions co-exist with serpentine motions for hot vapour condensates on nanotextured surface lubricated with silicone oil ($\eta$ = 2 mPa.s). Clockwise and anti-clockwise motions (indicated by full and dashed circles, respectively) appear with equal probability. Scale bar is 2 mm. (c) Clockwise circular motion for one droplet ($R$ = 0.5 mm) in (b).}
\end{figure*}




The polydispersity continues to increase, and now we have droplets of many different sizes ($R$ = 0.05--1 mm) performing serpentine motions [Fig.~\ref{fig:complex}(a) and Supporting Video S7]. The resulting breath figure exhibits self-similarity (at least over one order of magnitude), and the droplet motion looks roughly similar irrespective of their sizes (cf. inset and main image in Fig.~\ref{fig:complex}(a)). This self-similarity is a direct result of the lack of a natural length scale in the problem \cite{mandelbrot1982fractal}: $U_{\text{L}}$ is independent of $R$ and depends only indirectly on $\langle r \rangle$. In contrast, there is no self-similarity in jumping droplets where $U_{\text{V}} \propto 1/R^{1/2}$, and only small sub-millimetric but not large millimetric droplets jump off superhydrophic surfaces.  

As time progresses, the sweeping droplet motions \textit{locally} deplete the lubricant film \cite{kreder2018film} such that wetting ridges can no longer overlap to drive capillary attraction; in such lubricant-depleted regions, most of the droplets are immobile and are more uniform in size (compare left and right portions in Fig.~\ref{fig:complex}(b)). However, there are small pockets in the lubricant-depleted region with thicker lubricant overlayer where droplets ($R$ = 0.1--0.5 mm) remain mobile and move in circular motions, with both clockwise and anti-clockwise directions occuring with equal probability [Supporting Video S8]. Figure \ref{fig:complex}(c) is a close-up of one such (clockwise) circular motion. The droplet ($R$ = 0.5 mm) continually seeks areas with with higher $\phi_{l}$ but excluding lubricant depleted regions; the higher $\phi_{l}$ regions (indicated by asterisk) appear fuzzy because of the newly formed condensates (compare this to the clear trail left behind by the moving droplet where $\phi_{l} = 0$). The circular motion, which can be maintained over many cycles, is therefore similarly fuelled by the same release of interfacial energy and can be thought of as a special case of serpentine motion \textit{constrained} to areas with sufficient lubricant overlayer [Supporting Figs.~S9--S10].

These circular motions can co-exist with the self-similar serpentine motions on different areas of the same surface. As the lubricant is continually being redistributed across the substrate by the moving droplets, the various droplets continually switch between serpentine and circular motions in a highly collective fashion [Supporting Videos S6--S8]. The collective motion will eventually stop when $h_{o}$ becomes too thin for the wetting ridges of neighbouring droplets to overlap [Supporting Fig.~S10]. For Fig.~\ref{fig:serpentine_zoomed}(a)--\ref{fig:serpentine_zoomed}(b), droplet self-propulsion eventually stops after 250 min of continuous cooling [Supporting Fig.~S11 and Supporting Video S9]     

 The phenomenon described in this paper is general and applies to different substrates (micropillars vs. nanotextures) and experimental conditions (cooled substrate vs. hot vapour). The role of nanotextures/ micropillars is to retain the lubricant, and the propulsion speeds $U_{L}$ are insensitive to the exact texture geometries [Figs.~\ref{fig:serpentine_zoomed}(c)--~\ref{fig:serpentine_zoomed}(d)]. We observe the serpentine motion over a wide range of lubricant film thicknesses; practically, we found that an \textit{initial} $h_{o} = 5$ $\upmu$m for the micro-pillar and 15 $\upmu$m for the nanotextured surfaces are ideal to jumpstart the serpentine motion. Note, however, that once the serpentine motions start, there will be significant temporal and spatial variations in $h_{o}$ which drive the complex collective motion described in this paper, including the transition from serpentine to circular motions.      

To conclude, we have described how interfacial energy release can fuel the collective motion of condensate droplets on lubricated surfaces---an interesting example of emergent behaviour in active matter driven by condensation. Different modes of droplet motion from self-similar serpentine to circular motions were observed, and we propose a simple physical model to account for the observed droplet speeds based on energy balance. Our work has potentially important heat-transfer and water-harvesting applications.    

We would like to thank L. Mahadevan, A. Carlson, and J.V.I. Timonen for insightful discussions. This work was supported by KAUST start-up fund BAS/1/1416-01-01 and the Harvard Materials Research Science and Engineering Center under NSF Grant DMR-2011754.

\end{document}